\documentstyle[twocolumn,amssymb,aps,psfig,float]{revtex}

\begin{document}
\tolerance 50000

\draft

\twocolumn[\hsize\textwidth\columnwidth\hsize\csname @twocolumnfalse\endcsname

\title{Confinement and transverse conductivity in coupled Luttinger liquids}

\author{S. Capponi, D. Poilblanc and F. Mila
}
\address{
Laboratoire de Physique Quantique, \\ 
Universit\'e Paul Sabatier, 31062 Toulouse, France.
}

\date{June 96}
\maketitle

\begin{abstract}
\begin{center}
\parbox{14cm}{
One--particle interchain hopping in a system of coupled Luttinger liquids is 
investigated by use of exact diagonalizations techniques. Firstly, the two
chains problem of spinless fermions is studied in order to see the behaviour
of the band splitting as a function of the exponent $\alpha$ which
characterizes the $1D$ Luttinger liquid. 
Moderate intra-chain interactions can lead to a strong 
reduction of this splitting. The on-set of the 
confinement within the individual chains (defined by a vanishing splitting)
seems to be governed by $\alpha$.
We give numerical evidence that inter-chain coherent hopping can be 
totally suppressed for $\alpha\sim 0.4$ or even smaller 
$\alpha$ values.
The transverse conductivty is shown to exhibit a strong incoherent part.
Even when coherent inter-chain hopping is believed to occur
(at small $\alpha$ values),
it is shown that the coherent Drude weight is always significantly 
smaller than the incoherent weight. Implications for the optical experiments in
quasi-1D organic or high-$T_c$ superconductors is outlined. 
}
\end{center}
\end{abstract}

\pacs{
\hspace{1.9cm}
PACS numbers: 74.72.-h, 71.27.+a}
\vskip2pc]

Recently, the study of strongly correlated fermions confined to coupled
chains has received a great deal of interest in particular as a way of
studying the dimensional cross-over from 1D Luttinger-like behaviour to
2D. 

Some time ago, Anderson emphasized the crucial difference between
 in-plane and inter-plane (c-axis) transport observed in copper oxide
superconductors~\cite{Anderson_91}. Indeed, experimentally the
c-axis transport has an 
anomalous behaviour~\cite{Cooper} in the sense that the transverse conductivity has a
completely incoherent frequency dependence, there seems to be no sizeable
Drude-like term (except in the optimally doped systems) and $\sigma(\omega)$
is a very slowly increasing function of the frequency.
This phenomenon has been interpreted as an
incoherent hopping or as the ``confinement'' of 
the electrons inside the weakly coupled planes. 
However, for coupled Fermi liquids (FL), Landau theory predicts 
coherent transverse hopping and no anomalous transport. 
Therefore, it has been suggested that anomalous transport in the
direction of low conductivity is precisely the signature that the ground 
state (GS) of the two-dimensional (2D) plane itself is not of the usual FL 
type. Various candidates for this state have been proposed, such as the
marginal Fermi liquid~\cite{Varma} or the Luttinger liquid (LL) which is
generic in one dimension (1D)~\cite{Haldane}. 

Although non-Fermi liquid (NFL)
behaviour is thought to occur in the planes of high temperature 
superconductors (HTSC), it has been impossible yet to prove the NFL nature 
in 2D except for unrealistic models.  However, as stated above,
it is well known that the generic features of correlated 1D electrons 
are not FL-like but rather those of a LL and 
the precise nature (asymptotic behaviour, exponents, etc...) of the system 
can be easily controlled.
In addition, quasi-1D systems are realized in nature, and the problem of coupled
chains is of direct relevance there. For instance, in the case of the organic
superconductors of the (TMTSF)$_2$X family~\cite{jerome}, 
also known as the Bechgaard salts,
the high temperature properties are believed to be essentially one-dimensional,
while the low-temperature behaviour is three-dimensional. This cross-over is
presumably responsible for the anomalies observed in the temperature dependence
of several quantities (such as the $2k_F$ contribution to the relaxation 
rate~\cite{wzietek}, the ratio of the 
perpendicular conductivity to the parallel one~\cite{jerome}, 
the plasma edge when the electric
field is polarized perpendicular to the chains~\cite{jerome},\ldots) 
as well as for the
insulating behaviour reported for (TMTSF)$_2$ClO$_4$ in
the presence of a strong enough magnetic field~\cite{behnia}. 
Although a lot of work has
already been devoted to that problem~\cite{behnia,Boies}, 
several aspects of this cross-over have 
to be better understood, in particular those concerned with the transport 
properties perpendicular to the chains.

Hence, from now on and for sake of simplicity, we shall only deal 
with weakly coupled chains.

The effect of single-particle transverse hopping has previously been studied from a 
renormalization 
group point of view~\cite{Boies}. The notions of interchain coherence or incoherence
have emerged from such an analysis. 
Let us recall here that a LL has a different
structure from FL: there are no quasi-particle like excitations but instead
collective modes (charge and spin) with different velocities which lead to
the so-called spin-charge separation; moreover, the density of states $n(k)$
has no step-like structure at the Fermi level but instead a power 
law singularity
$n(k)-n(k_F)\sim|k-k_F|^\alpha$ defining the parameter $\alpha$ which depends on
the intra-chain interaction. 
It turns out that
the hopping $t_\perp$ is a relevant perturbation when
$\alpha\le 1$ \cite{Boies}. 

However, it has been argued that relevance in
that sense was not necessarily a sufficient condition to cause coherent 
motion between
chains. This, e.g., can be seen from the following model~\cite{Anderson_94};
let a system of two separated chains be prepared at time $t=0$ with a
difference of $\Delta N$ particles between the two chains. Then, the 
interchain hopping is turned on and one considers the probability of 
the system of
remaining in its initial state, $P(t)$. 
Coherence or incoherence can then be defined as the presence or absence 
of oscillations in $P(t)$.

In Ref.~\cite{Anderson_94}, the authors restricted themselves 
to the short time 
behaviour of $P(t)$ which, they argued, can provide valuable informations on 
interchain coherence or incoherence.  They found two regimes which 
depend entirely on the value of $\alpha$:
the case $\alpha< 1/2$ exhibited coherent motion while $\alpha>1/2$ showed no
signal of coherence.
However, as has been shown numerically and argued in Ref.~\cite{Didier2}, 
the small time behaviour can not always distinguish between 
coherent or incoherent regimes. In addition, if the initial state is 
far from equilibrium, i.e. if $\Delta N/ N$ is not infinitesimal,
then mathematical properties of the many-body spectrum related to 
the integrability or non integrability of the model will play a central role.
Indeed, in this case, coherent behaviour is found~\cite{Didier2} to be a generic
property of integrable models (irrespective of the value of $\alpha$). 
However, it is not clear whether this conclusion remains for low energy
excitations.

Another possible issue which has been raised is whether the spin degrees
of freedom are essential in Anderson's confinement mechanism or not~\cite{Clarke_96}.
The two chain problem with no anomalous exponent ($\alpha =0$) but
with spin-charge separation features was solved exactly in~\cite{Fabrizio} and showed
no confinement within each chain i.e. an energy separation was found between
the bonding and anti-bonding states. 
This result {\it a priori} does not contradict 
Anderson's conjecture which claims that the behaviour only depends on $\alpha$.
However, it is still not clear whether confinement can be found for spinless
models with large $\alpha$'s and whether the fact that the low energy 
excitations of the spinless fermion chain are collective modes is sufficient
to, alone, produce confinement in the chains. 

This paper is devoted to the study of various aspects of interchain coherence
in systems of strongly correlated spinless fermions. 
We shall derive several quantities sensitive to the
coherent/incoherent nature of the hopping transverse to the chains from
exact diagonalizations of small systems by the Lanczos
algorithm~\cite{Lanczos}. After a short presentation 
of the models in Sec. I, a precise characterization of the single chain 
system in terms of its LL parameters is given in Sec. II for models with 
screened interactions extending in space up to third neighbours.
In Sec. III, we study the simplest case of two coupled chains.
The influence of the interchain hopping on the splitting in energy of 
the singularities appearing in the spectral function for $k_\perp=0$ and 
$\pi$ is studied, in a similar way as has been done for
particles with spin~\cite{Didier}. Using different models with different 
extensions in space of the interaction, we investigate whether 
the physics depends only on the LL parameter $\alpha$ or not. 
In order to establish possible connections between the splitting 
and transport properties, we consider in Sec. IV a system of three
coupled chains. Periodic boundary conditions in the transverse
direction are used in order to derive the optical conductivity 
in this direction, hence, providing a direct (and experimentally 
observable) test of the confinement of the fermions within the chains.

\section{The model}
We consider here a model of spinless fermions on a lattice formed
by $m$ chains of length $L$ with a weak
inter-chain hopping:
\begin{eqnarray}
H=-\sum_{j,\beta} (c^\dagger_{j+1,\beta} c^{\phantom{\dagger}}_{j,\beta} + {\rm H.c.})\cr
 -t_\perp \sum_{j,\beta} (c^\dagger_{j,\beta+1} c^{\phantom{\dagger}}_{j,\beta}+ {\rm H.c.})\cr
 +\sum_{j,\beta,\delta} V(\delta)\, n_{j,\beta}\, n_{j+\delta,\beta}
\end{eqnarray}
where $\beta$ labels the chain ($\beta=1,\dots,m$), $j$ is a rung index
($j=1,\dots,L$), 
 $c_{j,\beta}$ is the fermionic operator which destroys one fermion at
site $j$ on the chain $\beta$, and $V(\delta)$ is a repulsive interaction between two fermions at a
distance $\delta$ (the lattice spacing has been set to one).  For
convenience, we choose a
repulsive interaction of the form $V(i)=2V/(i+1)$ for $i\le i_0$, with, more
specifically, $i_0=1,2,3$ which corresponds to an interaction extending up
to first, second and third nearest neighbours (NN) respectively.

In the x and y-directions, we shall use arbitrary boundary conditions (BC) by threading the system
with a magnetic flux $\Phi_x$ and $\Phi_y$ respectively (except for $m=1$ or 2
chains where open BC are used in the y-direction).
This is realized by a Peierls substitution that  modifies the kinetic term,
e.g. a twist along the transverse direction is realized by
\begin{equation}
c^\dagger_{j,\beta+1} c^{\phantom{\dagger}}_{j,\beta} \rightarrow c^\dagger_{j,\beta+1} c^{\phantom{\dagger}}_{j,\beta}
\,e^{i\frac{2\pi}{m}\Phi_y}
\end{equation}
where $\Phi_y$ is the flux measured in unit of the flux quantum
$\Phi_0=hc/e$. Similarly, the hopping term in the x-direction contains
a phase $\frac{2\pi}{L}\Phi_x$. 

The motivation to introduce flux is two-fold. Firstly, as proposed by 
Kohn~\cite{Kohn} transport properties of a correlated system can be 
directly measured from the response of the system to a twist in the boundary
condition. Secondly, our ultimate goal is to extract quantities in the 
thermodynamic limit from
finite size scaling analysis. It turns out that a simple way to improve the 
accuracy for a fixed system size is to average over the boundary conditions 
e.g. over $\Phi_y$ and/or $\Phi_x$~\cite{flux,Gros}.

\section{LL parameters for the single chain}
\begin{figure}[htb]
\begin{center}
\psfig{figure=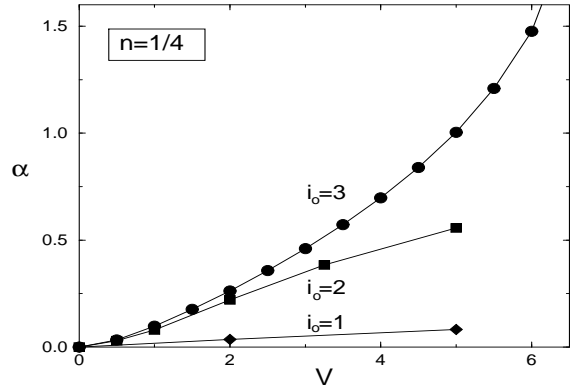,width=\columnwidth,height=6cm,angle=0}
\end{center}
\caption{Exponent $\alpha$ vs $V$ for $n=1/4$ and 
for NN interaction ($\blacklozenge$) 
and longer range $i_0=2$ ($\blacksquare$), $i_0=3$ ({\large$\bullet$}).}
\label{alpha_1D}
\end{figure}

In order to characterize the behaviour of coupled chains, it is required to,
first, compute the parameters of a single chain for the same model.
Indeed, one important issue is to study whether interchain transport is
a universal function of the LL parameters only or whether it depends on the 
details of the model. 
In the case of NN interactions, the hamiltonian (known as the t--V model)
can be mapped onto a spin chain problem by a Jordan-Wigner transformation
and this is exactly soluble by the Bethe ansatz; thus, $\alpha$ is known for 
each filling~\cite{Haldane}. However, for extended interactions in space, 
a numerical investigation is necessary with the help of conformal invariance
identities. It turns out that the exponent $\alpha$ 
can be related to simple physical quantities 
which can be easily extracted from standard exact diagonalization results
using the Lanczos algorithm. 
For instance the Drude weight $2\pi D$ ($D$ is the charge stiffness) and the 
charge velocity $u_\rho$ are related by~\cite{Voit}
\begin{equation}
\label{conformal}
2\pi D =2 u_\rho K_\rho ,
\end{equation}
where $K_\rho$ is a universal parameter which determines the 
long distance behaviour of the correlation functions.
The quantities $u_\rho$ and $D$ can be directly obtained on finite systems.  
The Drude weight for a single chain is given by the Kohn formula:
\begin{equation}
2\pi D=\frac{1}{4\pi}\frac{\partial^2 (E_0/L)}{\partial \Phi_x^2}
\end{equation}
The charge velocity can be extracted from the difference in energy of the
two singlet symmetry sectors $k=0$ and $k=2\pi /L$ 
(for an even number of fermions). 
$K_\rho$ is then obtained from Eq.~(\ref{conformal}) and 
the density of state exponent (for spinless fermions) 
can be calculated as~\cite{Voit}
\begin{equation}
\alpha=\frac{1}{2}(K_\rho+\frac{1}{K_\rho}-2)
\end{equation}
Finite size scaling analysis reveals that the $1/L^2$ law for the 
finite size corrections is very well satisfied for $K_\rho$ and
$\alpha$ so that an accurate determination of the extrapolated values 
can be calculated. Results for $\alpha$ are plotted in Fig.~\ref{alpha_1D}.

As is expected, the value of $\alpha$ increases with the strength of the 
interaction and with its range in space. However, for a given filling,
there is a maximum allowed value $\alpha_{max}$ for $\alpha$~\cite{Schulz}. 
Above this value, 
Umklapp processes become relevant, a gap opens up 
in the single particle spectrum and the system undergoes a
metal-insulator transition. The insulating phase is signalled by both an
opening of the single particle gap and by a
vanishing Drude weight~\cite{Kohn}. 
This region is of no interest to us in the present study since 
no transverse hopping occurs when $t_\perp$ is smaller than the 
single particle gap. For a given density, $\alpha$ acquires its maximum value 
on the metal-insulator transition line.
It is important to notice that this maximum value depends only on the 
density and, hence, the commensurability: if $n=p/q$, it is larger for 
larger $q$. As an example, for $n=1/2$
$\alpha_{max}=1/4$ while for $n=1/4$ a value as large as 
$\alpha_{max}=49/16\simeq 3.06$ can be obtained. 
This fact has motivated our choice of $n=1/4$ for the 
following calculations since it gives a large range of $\alpha$ values.
Typically, for $n=1/4$, $\alpha_{max}$ is realized by $i_0=3$
and $V\simeq 7.5$. For a shorter range interaction, significantly larger
values of $V$ are necessary.

\section{Interchain coherence}

The simplest approach to investigate interchain coherence is to 
consider 2 coupled chains, i.e. a $2\times L$ ladder.
We proceed along the lines of Ref.~\cite{Didier}. 
In the absence of interaction, $t_\perp$ leads to bonding and anti-bonding 
dispersion bands
corresponding to transverse momentum $k_\perp=0$ or $\pi$ respectively,
as seen in figure 2. 
The splitting $2t_\perp$ between these bands can be viewed as the signature of 
a coherent transverse hopping. 
These bands correspond to a $\delta$-function singularity
in the single particle hole (electron) spectral function 
for $k<k_F$ ($k>k_F$). 
The hole spectral function is defined by,
\begin{equation}
A_h ({\bf k},\omega)=-\frac{1}{\pi} 
Im\left\{\langle \phi_0|c^\dagger_{\bf k}
\frac{1}{\omega+i\varepsilon-H+E_0}
c^{\phantom{\dagger}}_{\bf k}|\phi_0 \rangle\right\}
\label{spectral_function}
\end{equation}
where ${\bf k}=(k,k_\perp)$ and a similar definition holds for 
$A_e$ by exchanging $c^{\dagger}_{\bf k}$ and $c^{\phantom{\dagger}}_{\bf k}$.

\begin{figure}[htb]
\begin{center}
\mbox{\psfig{figure=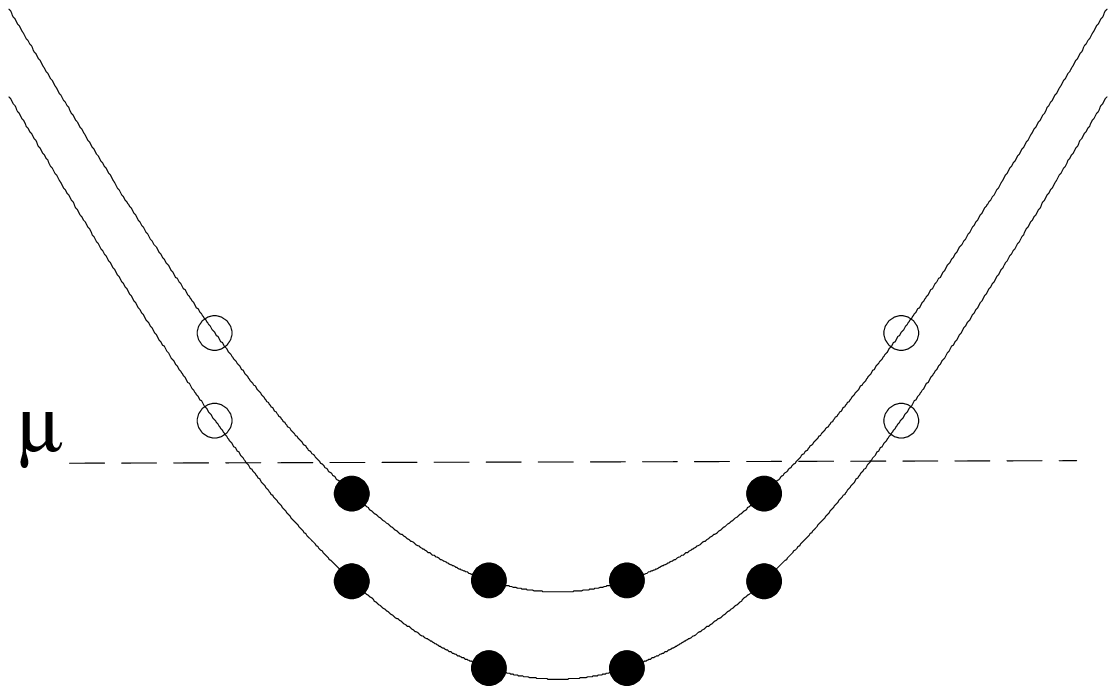,width=4.2cm,angle=0}}
\mbox{\psfig{figure=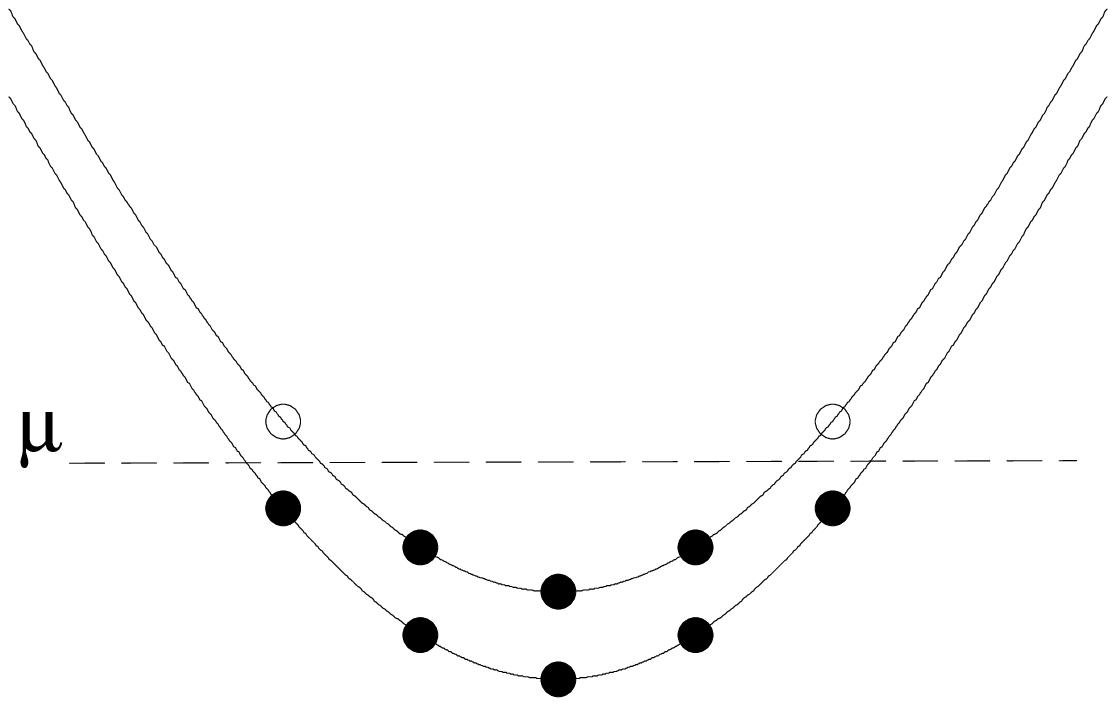,width=4.2cm,angle=0}}
\end{center}
\caption{Dispersion relation along the chain direction
in the closed (left) and open (right) shell configurations. Full (open)
symbols correspond to occupied (empty) states and 
$\mu$ is the chemical potential.}
\label{2chains}
\end{figure}

In the case of interacting particles, this $\delta$-function singularity
is replaced by a power law singularity and the elementary excitations
are collective modes. Here, we address the issue of the
influence of the hopping $t_\perp$ on this singularity, in particular we
investigate whether a splitting occurs. 
The choice of the boundary conditions in the x-direction is expected
to be important in the scaling behaviour of various quantities. 
Periodic or anti-periodic BC lead to 
closed or open shells as can be seen in figure~\ref{2chains}.
We shall consider these two cases separately. 

\subsection{Splitting: closed shell configuration}

In the closed shell configuration of Fig.~(\ref{2chains}),
we can add (respectively remove) one fermion on either of the two
branches (characterized by the transverse momentum $k_\perp$) 
just above (resp. below) the FL and then calculate the GS energy of
this new system. The difference between the two values of the energy gives
the splitting $\Delta E$ between the two bands.

For interacting particles, $\Delta E$ can be alternatively 
(and more precisely) 
defined as the energy separation between the two low energy peaks
in the spectral function $A_h({\bf k},\omega)$ and 
$A_e({\bf k},\omega)$ for $k_\perp=0$ and
$k_\perp=\pi$. For a longitudinal momentum k chosen above or
below the (non-interacting) Fermi wavevector $k_F=\pi n$ the electron
(hole) spectral function has to be considered. 
We have performed exact diagonalizations on clusters $2\times 4p$ with
$p=1,\dots,5$ at $n=1/4$ to obtain the splitting for different values of the
parameters. Results for a $2\times 16$ ladder at n=1/4 are shown in
Fig.~(\ref{spectral_func}) 
(a) and (b) for two momenta $k<k_F$ and $k>k_F$. The results look 
similar to the non-interacting case although the splitting between the 
$k_\perp=0$ and $k_\perp=\pi$ structures is reduced. Let us remark that a
new structure appears for $\omega <\mu$ and $k>k_F$. It is completely absent
in the non-interacting case but was predicted by Voit~\cite{Voit93} for large
$\alpha$'s by using a low energy bosonization scheme. Note, however, that
such calculations can not resolve the fine structure of the 
LL singularity.

\begin{figure}[htb]
\begin{center}
\psfig{figure=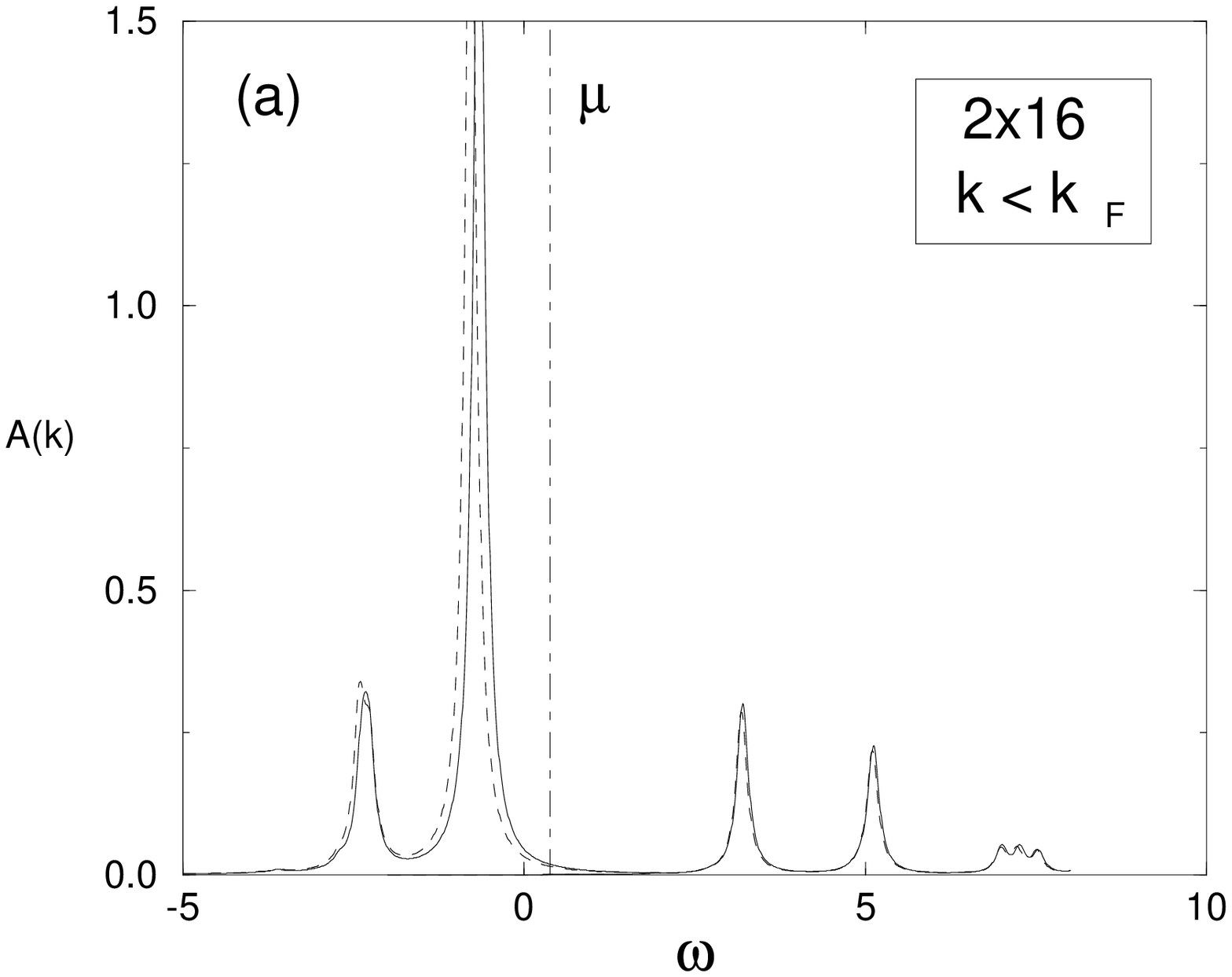,width=\columnwidth,angle=0}
\end{center}
\end{figure}

\begin{figure}[htb]
\begin{center}
\psfig{figure=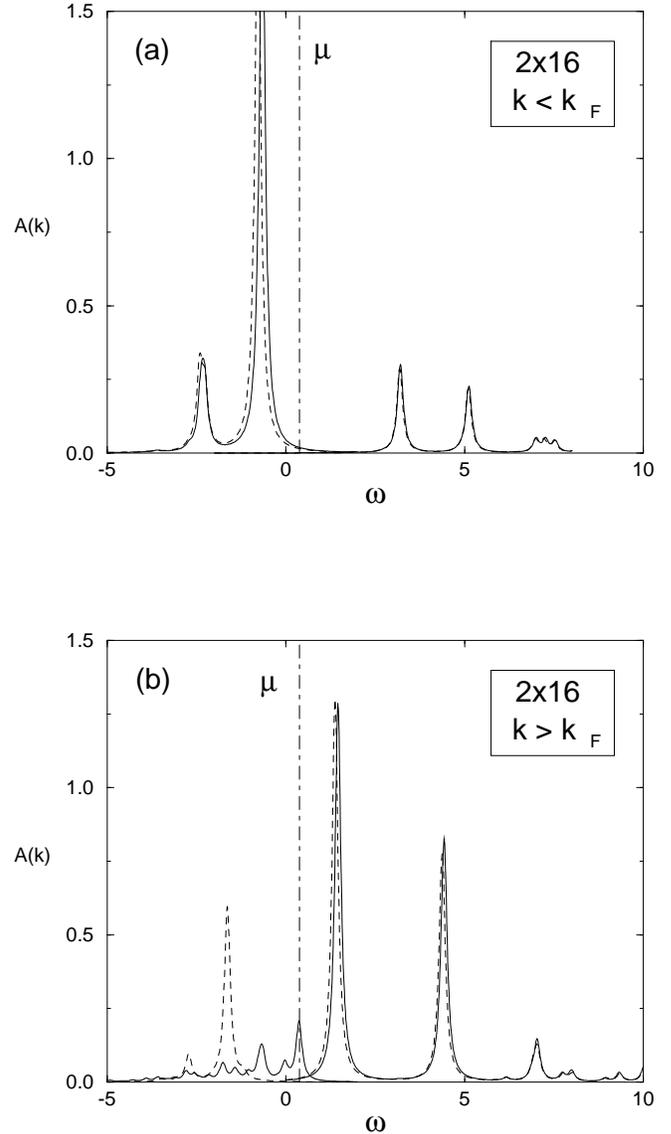,width=\columnwidth,angle=0}
\end{center}
\caption{Spectral function vs $\omega$ for the case $V=5$, $i_0=2$
with $t_\perp=0.1$. The regions $\omega<\mu$ and $\omega >\mu$
correspond to $A_h({\bf k},\omega)$ and $A_e({\bf k},\omega)$ respectively.
The full (dotted) lines correspond 
to $k_\perp=\pi$ ($k_\perp=0$).
(a) (resp. (b)) corresponds to a momentum below (resp. above) $k_F$. An
artificial width $\varepsilon$ has been added to the frequency,
$\omega\rightarrow \omega+i\varepsilon$.
}
\label{spectral_func}
\end{figure}

\begin{figure}[htb]
\begin{center}
\psfig{figure=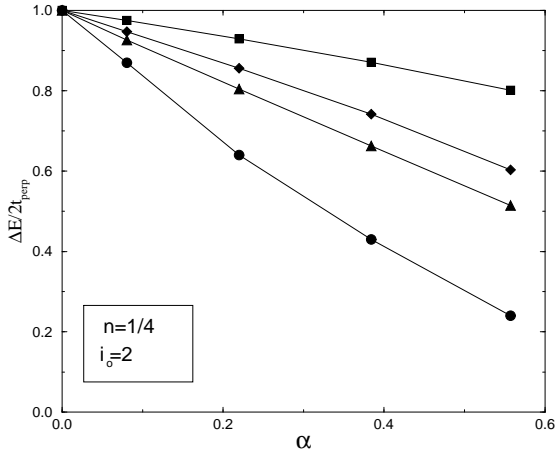,width=\columnwidth,angle=-90}
\end{center}
\caption{$\lim_{t_\perp \rightarrow 0}\Delta E/2t_\perp$ (=$a$) vs $\alpha$ 
for various system sizes and extrapolated values.
Closed shell configurations have been considered.
An average between the splitting above and below 
$k_F$ is performed. $L=8$, $L=12$ and  $L=16$ correspond to 
$\blacksquare$, $\blacklozenge$, $\blacktriangle$ respectively.}
\label{split_alpha.fig}
\end{figure}

 A careful study of the exact diagonalizations data reveals that $\Delta E$, for small $t_\perp$,
behaves like $a t_\perp +b t_\perp^3$ and the coefficient $a$ can be 
calculated accurately.
Finite size corrections of the order of $1/L$ have been found for $a$ 
so that $\lim_{t_\perp\rightarrow 0}\Delta E/2t_\perp$ can be estimated
accurately in the thermodynamic limit.
In Fig.~\ref{split_alpha.fig}, we plot, as a function 
of $\alpha$, the average between the splittings above and below the FL 
as well as the extrapolated value. 
The (normalized) splitting is greatly reduced by the electronic 
interactions (measured by $\alpha$) but it does not seem to vanish for
$\alpha=0.5$. On the contrary, it is compatible with the RG approach
predicting that $t_\perp$ becomes irrelevant, and hence that
$\lim_{t_\perp\rightarrow 0} \Delta E/2t_\perp$ vanishes, only for $\alpha>1$. 

However, these results must be considered with care for two reasons.
First, it is clear that $\Delta E$ alone does not completely characterize 
the transverse hopping dynamics. As seen in Ref. \cite{Didier},
the return probability $P(t)$, for example, depends on the whole 
frequency dependence of  $A_h({\bf k},\omega)$ 
(or $A_e({\bf k},\omega)$). Since, for large interactions, spectral 
weight from the LL singularity is transfered to higher excitation
energies, then an incoherent behaviour of $P(t)$ can be found even though
$\Delta E$ remains finite. 
Secondly, it should be stressed that the limit $t_\perp\rightarrow 0$ has been
taken first before eventually taking the limit 
$L\rightarrow\infty$. Indeed, the linear behavior 
of the splitting with $t_\perp$ is only valid in a narrow region which shrinks 
to zero as  $1/L$. In other words, when $t_\perp\sim \pi v_\rho / L$ ($v_\rho$
is some characteristic charge velocity), in the closed shell configuration,
the excitation energy at $k_\perp=\pi$ for a longitudinal momemtum just below 
$k_F$ crosses the chemical potential so that the splitting cannot any longer
be defined as some energy difference between many-body states corresponding
to the same number of particles. 
To avoid such complications, we now consider the open shell configuration
of Fig.~(\ref{2chains}).

\subsection{Splitting: open shell configuration}
\begin{figure}[htb]
\begin{center}
\psfig{figure=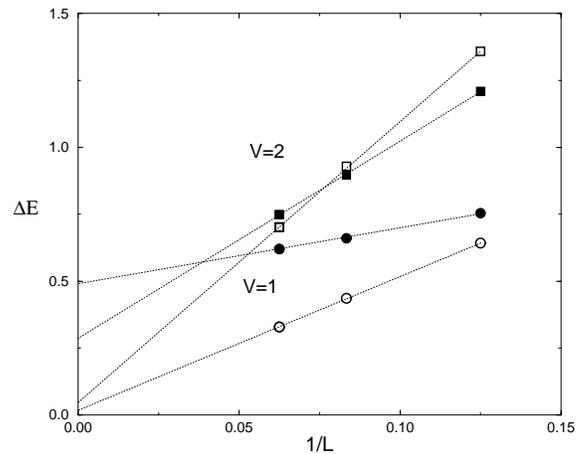,width=\columnwidth,angle=-90}
\end{center}
\caption{$\Delta E$ vs $1/L$ for $t_\perp=0$ (open symbols)
and $t_\perp=0.3$ (solid symbols).
Interactions up to next nearest neighbors ($i_0=2$) with different
strength V have been considered.}
\label{Splitting_scaling.fig}
\end{figure}

As can be seen in Fig. (\ref{2chains}), in the open shell configuration,
one can add or remove a particle exactly at the Fermi momentum
which is independent of the system size. 
We thus define the splitting as the difference between 
the electron and hole excitation energies i.e. 
$E_0(N_e+1,{\bf k}_e)-E_0(N_e)$ and 
$E_0(N_e)-E_0(N_e-1,{\bf k}_h)$ where $E_0(N_e)$ is the
reference energy corresponding to the absolute GS of the $N_e=nN$ electron
system. The momenta for the electron and hole excitations are fixed,
${\bf k}_e=(k_F,\pi)$ and ${\bf k}_h=(k_F,0)$.
Note that, for a fixed value of $t_\perp$, these elementary excitations 
are not the lowest energy excitations in the thermodynamic limit 
(if $t_\perp$ is relevant) since they
have a longitudinal momentum $k_F$ which is defined with respect to
the single chain case (i.e. for
$t_\perp=0$). 
We then naturally define,
\begin{equation}
\Delta E=E_0(N_e+1,{\bf k}_e)+E_0(N_e-1,{\bf k}_h)-2E_0(N_e).
\end{equation}
For V=0, this expression exactly gives the splitting $2t_\perp$ 
for any system size. For $t_\perp=0$ but finite interaction strength,
$\Delta E$ is finite. However, Fig.~(\ref{Splitting_scaling.fig}) 
reveals that it scales to zero in the thermodynamic limit as expected.
Fig.~(\ref{Splitting_scaling.fig}) 
also shows that an accurate finite size scaling analysis can be
performed for finite interaction strength and finite $t_\perp$, assuming 
1/L finite size corrections. 

\begin{figure}[htb]
\begin{center}
\psfig{figure=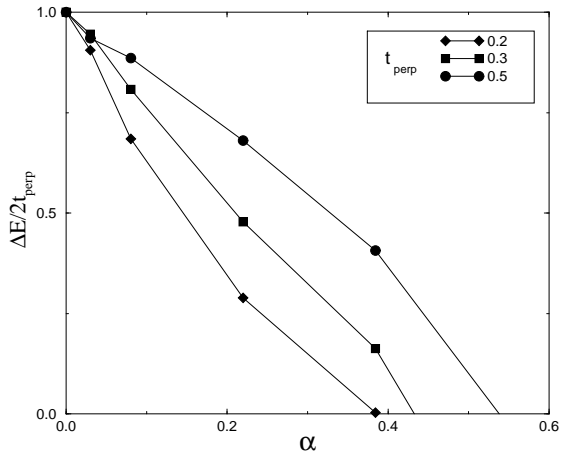,width=\columnwidth,angle=-90}
\end{center}
\caption{Extrapolated ($L=\infty$) values of $\Delta E/2t_\perp$ vs
$\alpha$ for several values of $t_\perp$ (indicated on the figure)
and in the $i_0=2$ spinless fermion model.
}
\label{Splitting_vsAlpha.fig}
\end{figure}

\begin{figure}[htb]
\begin{center}
\psfig{figure=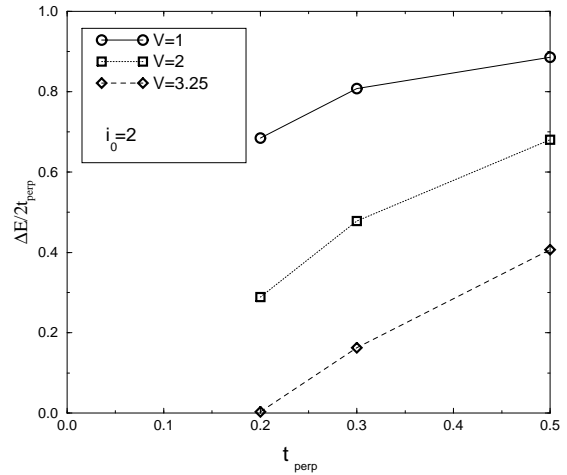,width=\columnwidth,angle=-90}
\end{center}
\caption{Extrapolated ($L=\infty$) values of $\Delta E/2t_\perp$ vs
$t_\perp$ in the $i_0=2$ spinless fermion model 
with several values of $V$ (indicated on the figure).
}
\label{Splitting_vsTperp.fig}
\end{figure}

The extraplolated values ($L=\infty$) of the ratio $\Delta/2t_\perp$
are plotted as a function of the exponent $\alpha$ of the 
$i_0=2$ model (i.e. with interactions extending up to second nearest 
neighbors) in Fig.~(\ref{Splitting_vsAlpha.fig}) for fixed interchain 
hopping amplitudes $t_\perp$.
A strong reduction of this ratio for increasing $\alpha$ indicates that
intrachain repulsion has a drastic influence in prohibiting
interchain coherent hopping. Our data suggest that there is a critical
value $\alpha^*(t_\perp)$ such that interchain hopping becomes
incoherent for $\alpha > \alpha^*(t_\perp)$. As expected, $\alpha^*(t_\perp)$ 
increases with $t_\perp$. Even for finite (but small) $t_\perp$,
$\alpha^*(t_\perp)$ can be as small as 0.4. 
The limiting value $\alpha^*(t_\perp=0)=\alpha_0$, although not
accurately given by our method, is probably significantly smaller than
0.4. This is indeed clear from Fig.~(\ref{Splitting_vsTperp.fig}) where
the same ratio $\Delta/2t_\perp$ is plotted as a function of $t_\perp$
for various interactions ($i_0=2$ model). For large interactions 
like $V=3.25$ ($\alpha\sim 0.38$), there is a critical value 
$t_\perp^*(\alpha)$ of $t_\perp$ below which incoherent transverse
hopping takes place. This corresponds to the case $\alpha > \alpha_0$.
For smaller interaction like $V=1$ ($\alpha\sim 0.08$) our data are
consistent with $\Delta/2t_\perp$ approaching a finite value when
$t_\perp\rightarrow 0$. Hence, transverse hopping remains coherent in
this case which corresponds to $\alpha < \alpha_0$.
For intermediate interactions like $V=2$ ($\alpha\sim 0.22$),
our data are not conclusive. However, very small values of 
$\alpha_0$ like 0.2 (or even smaller) 
are not inconsistent with our numerical analysis. 

It is important to notice that the previous extrapolations carried
out for open shell configurations and finite $t_\perp$ give very 
different results than those shown in Fig.~(\ref{split_alpha.fig}). 
In particular, Fig.~(\ref{Splitting_vsAlpha.fig}) suggests that 
the critical value $\alpha_0$ is significantly smaller than the value of
1 predicted by the RG analysis. 
One possible explaination is that the limits 
$t_\perp\rightarrow 0$ and $L\rightarrow\infty$ do not
commute with each other. This scenario is supported by the fact that,
in the closed shell configuration, the linear dependance of the 
splitting $\Delta E$ with $t_\perp$ is limited, for incresing system
size, to a narrower and narrower range of order $1/L$. 
Therefore, an extrapolation at finite $t_\perp$ has to be realized. 
It is interesting to note that the RG approach
seems, on the contrary, to reproduce the
$\lim_{L\rightarrow\infty}\{ \lim_{t_\perp\rightarrow 0} 
(\Delta E/ 2t_\perp)\} $ data.

We shall finish this section with a brief comment on the role of the 
spin degrees of freedom which have not been considered here.
It is clear that, when spin is taken into account, 
the spin-charge separation that
occurs in 1D should suppress even further the coherent transverse hopping.
At a qualitative level, this can be understood from the fact that only 
real electrons can hop from one chain to the next
and this is believed to become more difficult in the presence of spin-charge
separation as has been suggested by Anderson~\cite{Anderson_91}.
Although dealing with a different filling $n=1/3$ and with particles with
spin, the same qualitative behaviour is found numerically in
Ref.~\cite{Didier}; however, the decrease of the splitting with $\alpha$ is
stronger.

\section{Transport properties}

The previous study suggests that the interaction tends to confine the 
electrons within the chains, although no complete confinement 
seems to occur at small $\alpha$ values where $\Delta E\ne 0$. 
A better understanding of this phenomenon can be achieved by investigating
the transport properties along 
the y-axis (inter chain) and more precisely the transverse optical
conductivity which is the linear response of the system to a spatially
uniform, time dependent electric field in the transverse direction.
For such a study, a torus geometry is needed ($m\ge 3$) so that a current
can flow around the loop in the y-direction. 
One of the main advantage of the optical conductivity 
is that it can directly be measured experimentally. 

The real part of the optical conductivity can be written as a sum of two parts,
\begin{equation}
 \sigma_{yy}(\omega)=2\pi D_{yy} \,\delta(\omega)+\sigma_{yy}^{reg}(\omega)
\end{equation}
where $D_{yy}$ is the charge stiffness in the y-direction which 
defines the Drude weight $2\pi D_{yy}$. 
Note that an important 
f-sum rule~\cite{Maldague} can be used to check the numerical 
results,
\begin{eqnarray}
\label{sum}
\int_0^{\infty} d\omega\,\sigma_{yy}(\omega)&=&\frac{\pi e^2}{2N}\langle
t_\perp \sum_{j,\beta} (c^\dagger_{j,\beta+1} c_{j,\beta}+ {\rm H.c.})\rangle
\nonumber \\
&=&-\frac{\pi e^2}{2N} T_{yy}  ,
\end{eqnarray}
where the expression between brackets is the mean value of the 
transverse kinetic energy in the ground state. 

The actual calculation of the
frequency dependence of $\sigma_{yy}^{reg}(\omega)$ will be carried out
later on and we first concentrate on $D_{yy}$ and on the total sum rule.
As originally noted by Kohn~\cite{Kohn}, $D_{yy}$, which measures a transport
quantity, 
can be obtained from the dependence of the ground state energy $E_0$ on 
$\Phi_y$ as
\begin{equation}
\label{Kohn.eq}
2\pi D_{yy}(\Phi_y)=\frac{m^2}{4\pi}\frac{\partial^2 (E_0/N)}{\partial \Phi_y^2}
\end{equation}
where $N=mL$ is the number of sites. 
This is, in all points, very similar to the previous 
derivation of the charge stiffness $D$ for the 1D rings. 

\begin{figure}[htb]
\begin{center}
\psfig{figure=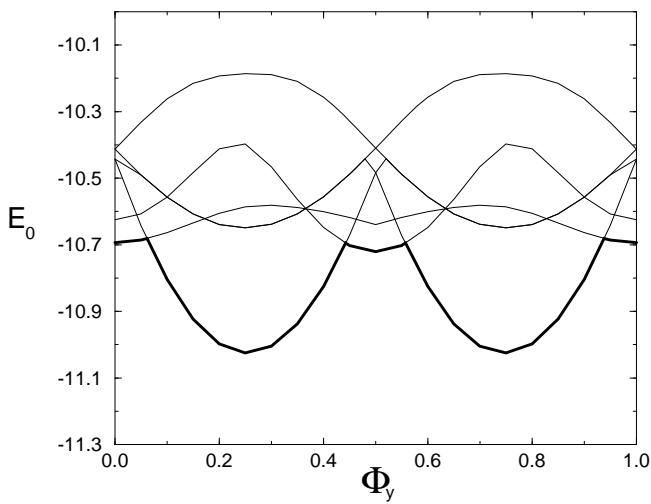,width=\columnwidth,angle=0}
\end{center}
\caption{ GS energy vs $\Phi_y$ for a $3\times 8$ systems with $V=2$,
$i_0=2$, $t_\perp=0.7$ and closed shell configuration at quarter filling . The thick curve is the absolute GS energy.
}
\label{EvsPhi}
\end{figure}

The previous quantities have been calculated numerically on finite $3\times L$
lattices using the Lanczos algorithm. 
We have chosen a quarter filled band so that
the extrapolation of results for $L=4,8$ and $12$ is possible. From now on, we shall restrict ourselves 
to PBC or ABC in the x-direction.
In most cases, an average over $\Phi_y$ is realized~\cite{flux,Gros}
in order to mimic the case of many parallel chains
i.e. we calculate,
\begin{equation}
\langle D_{yy} \rangle_{\Phi_y}  =\int_{-1/2}^{1/2} d\Phi_y\, D_{yy}(\Phi_y).
\label{average}
\end{equation}
This is done by calculating the ground state energy for a (large) discrete
set of flux values (still making use of the translational 
invariance). A typical curve for $E_0(\Phi_y)$ is shown in Fig.~(\ref{EvsPhi})
and reveals that level crossings occur as a function of $\Phi_y$. 
As for the ladder case, the scaling behaviour depends crucially 
on the choice of the boundary conditions along the x-direction.
In the following, we shall study two different cases separately.

\subsection{Charge stiffness: closed shell configuration}

\begin{figure}[htb]
\begin{center}
\mbox{\psfig{figure=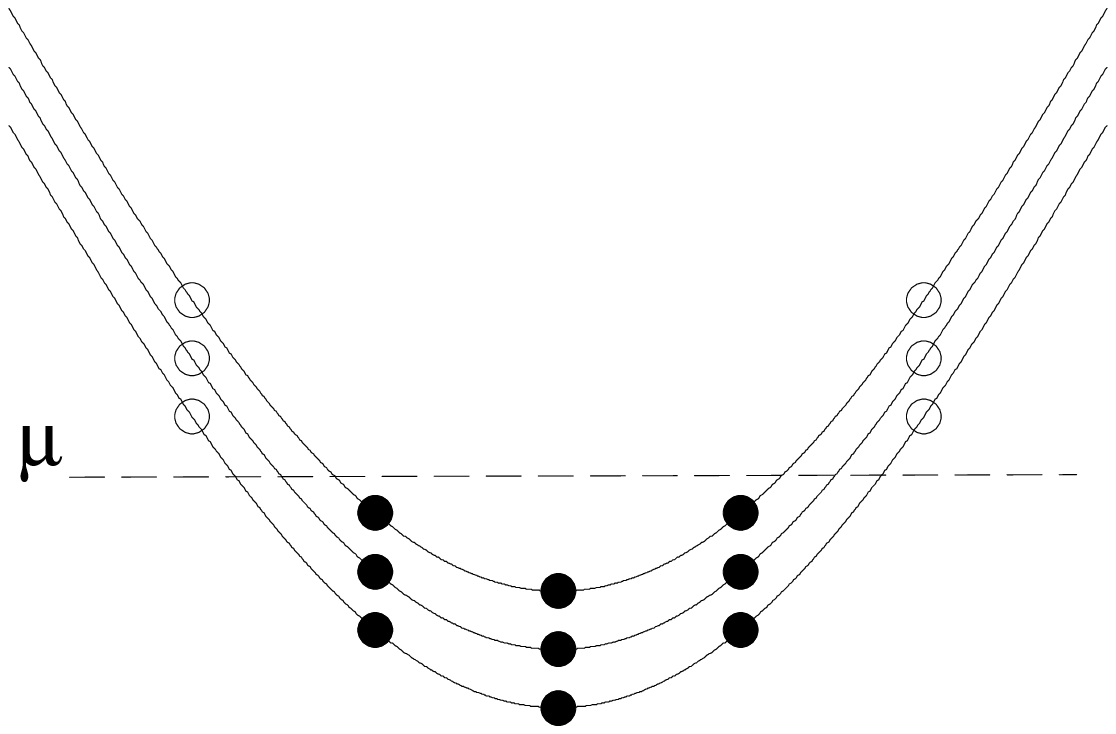,width=4.2cm,angle=0,clip=}}
\mbox{\psfig{figure=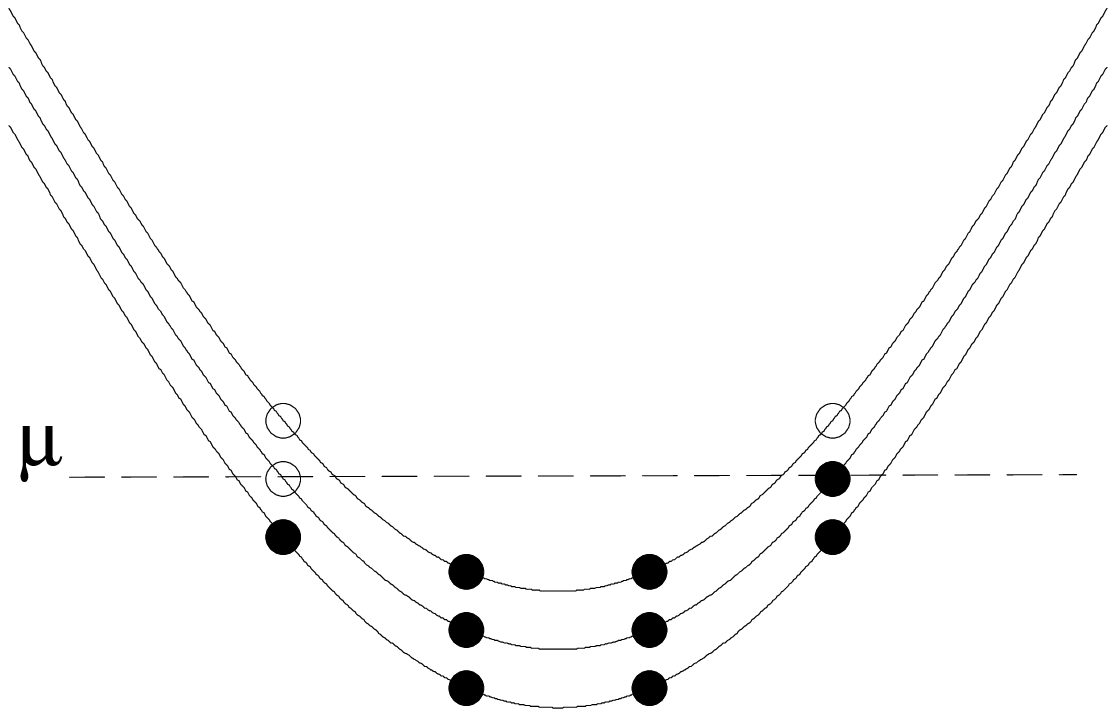,width=4.2cm,angle=0,clip=}}
\end{center}
\caption{Dispersion relations as a function of $k$ for closed and open shells.}
\label{3_chains}
\end{figure}

In the non-interacting case of a $3\times L$ system with 
periodic BC in both directions, $t_\perp$ leads to
three branches in $k$-space separated by an energy of order  
$\sim t_\perp$ (see Fig.~\ref{3_chains}). Note
that for special values of $\Phi_y$ (e.g. $\Phi_y=0,\frac{1}{2}$), 
two of them are degenerate.
We first consider the closed shell configuration  
where, for a given momentum k along the chain direction (and sufficiently 
small $t_\perp$), the three possible 
states with different $k_\perp$ momenta ($0$, $\frac{2\pi}{3}$ or 
$-\frac{2\pi}{3}$) are either fully occupied or completely empty.
In this case, there is a critical value $t_\perp^*$ of $t^{\phantom{*}}_\perp$
for which, for an
optimum choice of $\Phi_y$, one can move a fermion from one branch to the next 
with no energy cost. A straightforward
calculation gives 
$t_\perp^*=\sqrt{2/3}\sin(\pi/L)$.
Below this value, in the y-direction the bands are filled and therefore 
no transport can occur in the
transverse direction and the Drude weight is vanishing. Note that this
quantity $t_\perp^*$ is directly related to the splitting of the bands.

\begin{figure}[htb]
\begin{center}
\psfig{figure=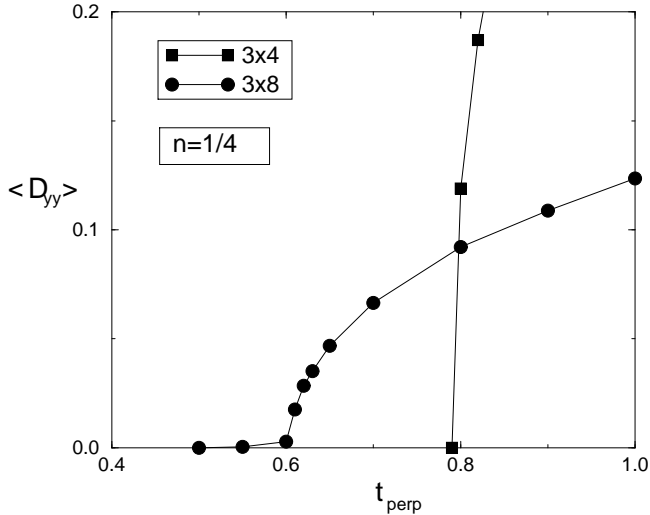,width=\columnwidth,angle=0, clip=}
\end{center}
\caption{Drude weight in arbitrary units as a function of $t_\perp$ for the
case $V=2, i_0=3$ on the $3\times 4$ and $3\times 8$ lattices. An average over
$\Phi_y$ has been performed according to Eq.~(\protect\ref{average}). }
\label{D_close.fig}
\end{figure}

More generally in the interacting case, the same behaviour is observed as
can be seen in Fig.~\ref{D_close.fig} where we plot the Drude weight 
as a function of $t_\perp$ for a given size. 
This is a typical behaviour from which one can define some precise 
cross-over value $t_\perp^*(L)$
where a transition occurs ($D_{yy}$ increases suddenly). 
Of course, this behaviour is a finite size effect
and we must be careful to extract thermodynamic results properly.

Two scenarios can occur. Firstly, $t_\perp^*(\infty)$ might remain 
finite. In this case, no coherent transport occurs between the chains for
sufficiently small $t_\perp$. This is the case, for example, for very strong
interactions in the insulating phase. This is also expected in the LL phase
for $\alpha>\alpha_0$. 
Indeed, we expect that $\Delta E=0$ (see sec. III) 
would imply $D_{yy}=0$ (for $L=\infty$).
Although it is difficult to prove this behaviour
numerically, it is not incompatible with our numerical results. 
The second possibility is that $t_\perp^*(L)\rightarrow 0$ when 
$L\rightarrow \infty$. In this case, we expect that the non-interacting 
picture should be approximately valid, at least for not too large 
an interaction.
In other words, according to the non-interacting picture which we discussed
above, we expect $t_\perp^*(L)$ to be directly related to the
splitting $\Delta E$ calculated in the previous section. 

\begin{figure}[htb]
\begin{center}
\psfig{figure=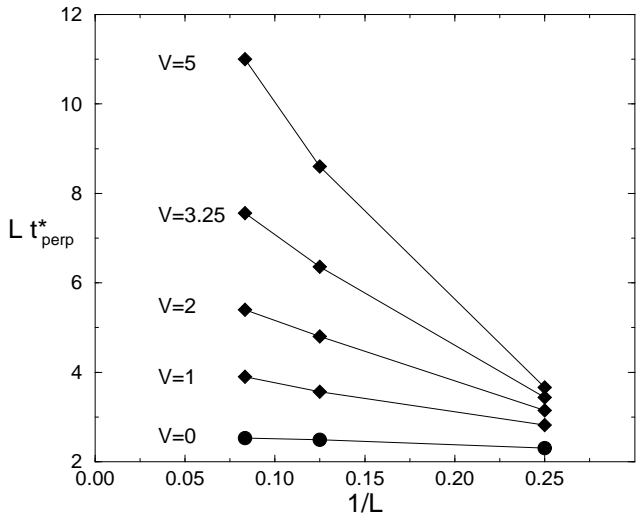,width=\columnwidth,angle=0}
\end{center}
\caption{$L t_\perp^*(L)$ as a function of $1/L$ for the non-interacting
case ({\large $\bullet$}) and for finite  interaction $i_0=2$ and
$V=1,2,3.25,5$ ($\blacklozenge$)}
\label{Ltperp}
\end{figure}

\begin{figure}[htb]
\begin{center}
\psfig{figure=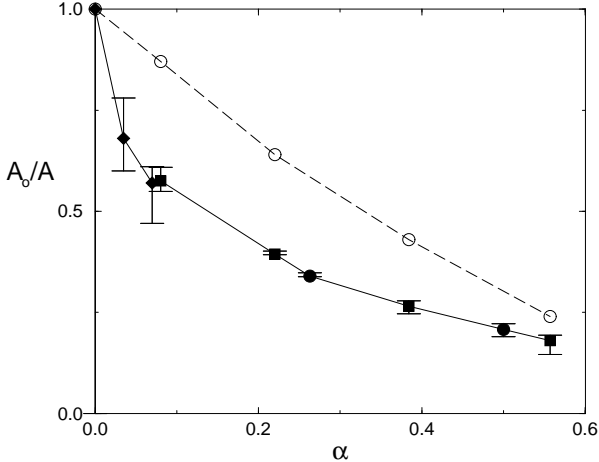,width=\columnwidth,angle=0}
\end{center}
\caption{
Extrapolated values of $A_0/A$ as a function of $\alpha$ 
for different range of interactions $i_0=1$ ($\blacklozenge$), $i_0=2$ 
($\blacksquare$) and  $i_0=3$ ({\large $\bullet$}).
For comparison, the extrapolated values of 
$\Delta E$ of Fig.~(\protect\ref{split_alpha.fig}) are also reproduced 
here (open circles).
}
\label{tperp_alpha.fig}
\end{figure}

We proceed as follows; we assume a connection 
between $t_\perp^*(L)$ and the splitting $\Delta E$ and show that this
hypothesis is, indeed, consistent with the numerical data.
Let us suppose that $t_\perp^*(L)$ fulfills,
\begin{equation}
\max_{\Phi_y}{ \{\Delta E(t_\perp^*,\Phi_y)\} }\sim\frac{2\pi}{L} v_\rho,
\label{argument}
\end{equation}
where $v_\rho$ is some charge velocity. 
Thus, $t_\perp^*(L)$ should be of order $1/L$,
more precisely it should behave like $t_\perp^*(L)=A/L+B/L^2$. 
Finite-size scaling can be performed by plotting $L t_\perp^*(L)$ vs 
$1/L$. As seen in Fig.~(\ref{Ltperp}), this scaling law seems to be
well satisfied for small interaction strength so that, in this case,
a finite-size extrapolation of $A$ is possible. On the other hand,
a diverging value of $Lt_\perp$ (or A)
would mean that $t_\perp^*(\infty)$ is in fact finite and, hence, completely
incoherent hopping occurs. 
This is likely to occur for the largest values of $V$ we have considered, 
such as $V=3.25$ and $V=5$ and it can not be completely excluded 
for smaller values of $V$.
According to the qualitative argument of Eq.~(\ref{argument}), 
if $\Delta E(t_\perp^*,\Phi_y)$ is linear in $t_\perp$ for small $t_\perp$,
then $1/A$ is expected to be directly proportional to the extrapolated value
of $\Delta E$. 
In order to check this point, it is convenient to normalize $A$ with 
respect to the non-interacting case, $A_0=\sqrt{2/3}\,\pi$ for $n=1/4$.
The quantity $A/A_0$ is plotted
as a function of the Luttinger parameter $\alpha$ for different values of
the interaction in Fig.~\ref{tperp_alpha.fig} and compared to the extrapolated
value of $\Delta E$ obtained for closed shell configurations. 
Firstly, we remark that, although some error bars are large,
the behaviour of $t_\perp^*$ seems to be quite similar for different kinds of
interactions. In other words, $\alpha$ seems to be the only 
parameter controlling the behaviour of $1/A$.
Secondly, we observe some discrepancies between $1/A$ and $\Delta E$.
They can be attributed (i) to the crudeness of
the picture we have developed only at a qualitative level, (ii) to the 
strong dependence of the charge velocity  $v_\rho$ in
Eq.~(\ref{argument}) with the interaction and (iii) 
more importantly, to the fact that the
previous estimation  of the splitting realized in the closed shell
configuration case seems not to be relevant for finite $t_\perp$. Indeed,
the second estimation of Sec. III.B of the ratio $\Delta E/2t_\perp$,
realized for 
open shell configurations and finite $t_\perp$, gives, for small $\alpha$,
smaller values in better agreement with $A_0/A$. Moreover, 
the above extrapolation of $t^*_\perp (L)$ assumes $t_\perp^*(L=\infty)=0$,
i.e. coherent interchain transport. However, as mentioned above, the 
possibility of a {\it small} finite value of $t^*_\perp (L=\infty)$
cannot certainly be excluded when $\alpha > 0.2$ which would be in complete
agreement with the results for $\Delta E/2t_\perp$ obtained 
in the open shell case.

At this point, we are still left with an incomplete picture. 
For not too large an interaction ($\alpha < \alpha_0$), 
our results suggest that $D_{yy}$ is finite for small transverse hopping.
Fig.~(\ref{D_close.fig}) suggests that the slope
 ${\partial\langle D_{yy}\rangle}/{\partial t_\perp}
\mid_{t_\perp=t_\perp^*}$ seems to vanish when $L\rightarrow\infty$.
Therefore, a $t_\perp^2$ behavior of the Drude weight with $t_\perp$ 
is certainly possible as in the non-interacting case.
To have a better understanding of this behaviour we now 
use a different choice of the boundary conditions along x.


\subsection{Charge stiffness: open shell configuration}

\begin{figure}[htb]
\begin{center}
\psfig{figure=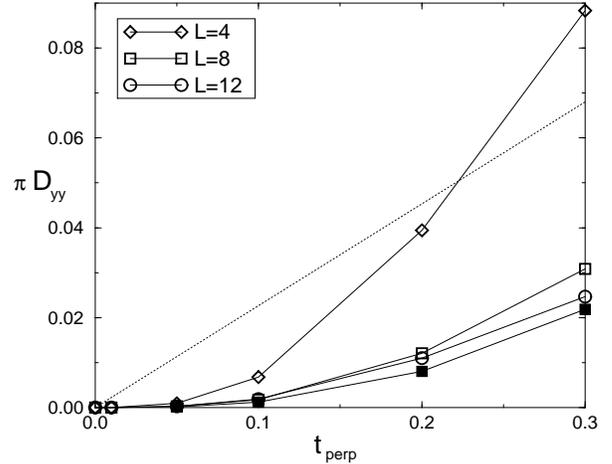,width=\columnwidth,angle=0}
\end{center}
\caption{$\pi D_{yy}$ for a fixed value of ${\Phi_y}$ (see text),
$V=2$, $i_0=3$ and different 
system sizes $L=4,8,12$ from up to down (open symbols) compared to the
averaged data of $L=8$ (filled squares). The non-interacting case 
(for a $3\times 8$ torus) is shown as a dotted line. 
}
\label{D_phi.fig}
\end{figure}

By choosing adequate boundary conditions along x ($\Phi_x=0$
or $\Phi_x=\pi$ depending
on the length $L$), open shells can be realized as seen in
Fig.~(\ref{3_chains}).
In this case, even for a finite system, the Drude weight and the total kinetic 
energy remains finite down to vanishing $t_\perp$ as in the
non-interacting case. We first consider a fixed value of the flux
$\Phi_y$. It is in fact convenient to choose $\Phi_y$ corresponding to the
lowest GS energy, i.e. such that $\frac{\partial E_0}{\partial \Phi_y} =0$,
since this value (in fact $\Phi_y\sim 0.25$) 
is almost independent of the interaction and of the system size. 
The corresponding Drude weight
is shown in Fig.~\ref{D_phi.fig} for $3\times 4$, $3\times 8$
and $3\times 12$ systems as a function of $t_\perp$. 
Finite size effects are found to be already weak 
for the two largest cases $L=8$ and $L=12$. The Drude weight is clearly 
strongly suppressed compared to the $V=0$ case~\cite{note_Convergence}. 

\begin{figure}[htb]
\begin{center}
\psfig{figure=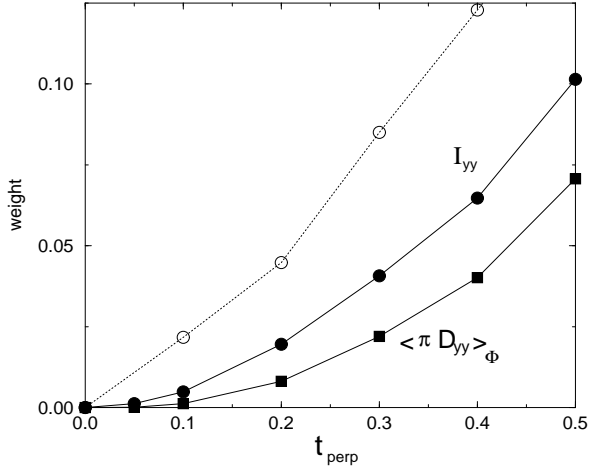,width=\columnwidth,angle=0}
\end{center}
\caption{
$\pi\langle D_{yy}\rangle_{\Phi_y}$ ($\blacksquare$) and $I_{yy}$
({\large $\bullet$}) as a function of $t_\perp$ for a
$3\times 8$ system and $V=2$, $i_0=3$.
For comparison, the $V=0$ case is also displayed as a dotted line.
}
\label{DT.fig}
\end{figure}

We now consider an average over $\Phi_y$ and 
we shall denote by
$I_{yy}=\langle \int_0^\infty \sigma_{yy}(\omega) d\omega 
\rangle_{\Phi_y}$ the total sum which is obtained by computing the 
transverse kinetic energy and averaging over $\Phi_y$. 
As seen in Fig.~\ref{D_phi.fig}, the values obtained for the 
Drude weight by averaging the $3\times 8$ data over $\Phi_y$ 
are very close to the ones obtained on the $3\times 12$ cluster 
at constant flux. Therefore, it is technically avantageous, as far as
CPU time is concerned, to consider smaller systems and perform a flux average.
Note that, however, if the number of coupled chains is kept fixed
(here $m=3$), even 
in the limit $L\rightarrow\infty$, we expect 
$\langle D_{yy}\rangle_{\Phi_y}\ne D_{yy}(\Phi_y)$ (since the
thermodynamic limit is not taken in the direction of the flux). 
Qualitatively, the averaging procedure mimics many coupled chains.
In any case, when confinement within each of the individual chains 
starts to occur we do not expect crutial differences between the cases
of 3 or of an infinite number of coupled chains (if only the $t_\perp$ term
couples the chains). 
In Fig.~\ref{DT.fig}, we plot the Drude weight and the total sum rule 
as a function of $t_\perp$ for a moderate electronic interaction.
We observe that the behavior of these two quantities is not 
incompatible with the $t_\perp^2$ law of the non-interacting 
case~\cite{note_Convergence}.
However, the intrachain interaction has drastic effects.
Firstly, the total sum rule is strongly reduced 
compared to $V=0$ (shown as a reference on Fig.~\ref{DT.fig}).
Secondly, it is found that $\pi\langle D_{yy}\rangle_{\Phi_y}$ and $I_{yy}$ 
behave differently. In other words, for small $t_\perp$ 
the main fraction of the weight lies in the incoherent part. 

To be more quantitative, it is convenient to consider the ratio 
\begin{equation}
r=\frac{\pi \langle D_{yy}\rangle_{\Phi_y}}{I_{yy}}
=\frac{m^2}{4\pi^2}\left\langle
\frac{\partial^2 E_0}{\partial \Phi_y^2} \right\rangle_{\Phi_y} /
\left\langle -{T}_{yy} \right\rangle_{\Phi_y}  ,
\label{ratio}
\end{equation}
which, as can be seen from the sum rule Eq.~(\ref{sum}),
corresponds to the relative part of the Drude weight
in the total optical conductivity. We have plotted the ratio 
$r$ for the same $3\times 8$ torus and for the 
three interaction ranges, $i_0=1$, $2$ and $3$.
The common important feature of these data is that the ratio $r$ decreases
as $t_\perp$ goes to $0$ and can become rather small, typically smaller
than $20\%$. Unfortunately, we think our data become somewhat unreliable
for very small $t_\perp$ (let's say $t_\perp<0.1$) so that the behavior
of the ratio r when $t_\perp\rightarrow 0$ cannot be accurately determined.
However, our numerical estimates should give the correct trend in the 
range $0.1<t_\perp <0.4$. 

\begin{figure}[htb]
\begin{center}
\psfig{figure=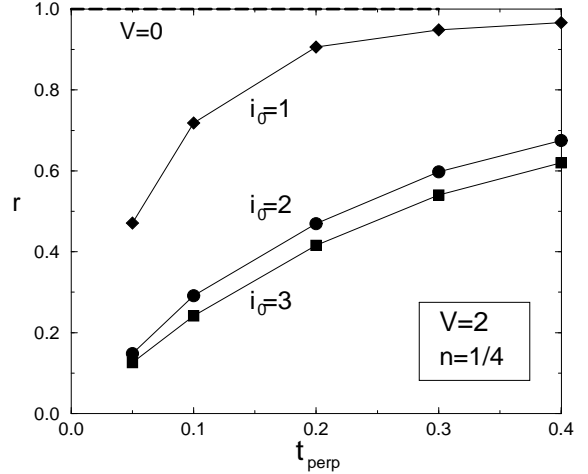,width=\columnwidth,angle=0}
\end{center}
\caption{Ratio of the Drude weight to the total conductivity as a function
of $t_\perp$ for NN and 
longer range interaction with the same NN magnitudes $V=2$.}
\end{figure}

These results should have very important consequences on the experimental 
side. Indeed, our results predict, for small $t_\perp$, anomalous transport
perpendicular to the chains even when $D_{yy}$ does not completely 
vanish (if $\alpha<\alpha_0$). Spectral weight is suppressed 
predominantly from the coherent part of the conductivity.
In order to confirm this behaviour, the frequency dependent optical
conductivity has been  calculated directly and results are 
discussed in the following.

\subsection{Optical conductivity}

The frequency dependence of the transverse optical conductivity 
can be calculated by use of the Kubo formula,
\begin{equation}
\sigma_{yy}^{reg}(\omega)=\frac{\pi}{N}\sum_{n\ne 0} \frac{|\langle
\phi_0|\,\hat{\j}_y\,|\phi_n\rangle |^2}{E_n-E_0}\, \delta(\omega-(E_n-E_0))
\end{equation}
where $\hat{\j}_y$ is the transverse current operator and the sum runs over
all the excited states.
So far, we restrict ourselves to the open shell configuration used in the preceding
Section. $\sigma_{yy}^{reg}(\omega)$ can be calculated exactly 
on the same finite clusters by 
a continuous fraction expansion generated by use of the Lanczos method.
Firstly, for a given value of $\Phi_y$, one computes the ground state.
Here, we choose the absolute GS which carries no current in the
y-direction. Then, by applying the transverse current operator on 
this state, one generates a new vector which serves as the starting 
point of another Lanczos procedure.
Eventually, the tridiagonal form of the hamiltonian in this new basis 
is used to compute the
continued fraction expansion of the regular part of $\sigma_{yy}(\omega)$.

\begin{figure}[htb]
\begin{center}
\psfig{figure=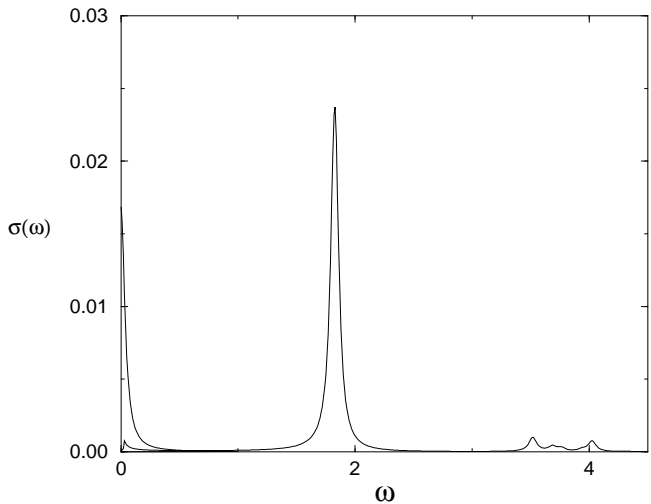,width=\columnwidth,angle=0}
\end{center}
\caption{Transverse optical conductivity vs frequency 
for a $3\times 8$ system at quarter filling with $V=2$, $i_0=3$ and $t_\perp=0.1$.
The Drude $\delta$-function has been represented with the
same small imaginary part $\varepsilon=0.04$.
}
\label{conductivity.fig}
\end{figure}

\begin{figure}[htb]
\begin{center}
\psfig{figure=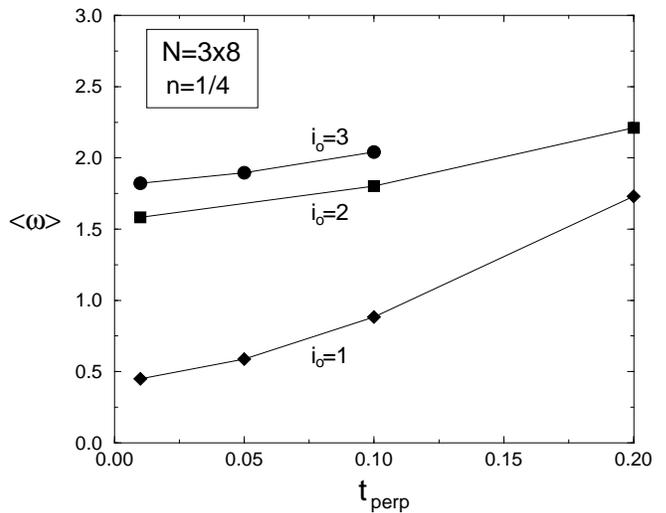,width=\columnwidth,angle=0}
\end{center}
\caption{First moment of the conductivity as a function of $t_\perp$
for $V=2$ and different interaction ranges.}
\label{moment}
\end{figure}

In the free case, the current operator commutes with the Hamiltonian and
therefore, the conductivity only contains a Drude peak.
However, as can be seen on Fig.~\ref{conductivity.fig},
 for finite interaction strengths and in
the $3\times 8$ cluster, a
pronounced structure appears at finite frequency for any small value of 
$t_\perp$. In order to determine more accurately the position in energy 
of the weight, we have computed the first moment of the distribution,
\begin{equation}
\langle \omega\rangle=\int_{0^+}^\infty
\sigma_{yy}^{reg}(\omega)\,\omega\,d\omega \;/ \int_{0^+}^\infty
\sigma_{yy}^{reg}(\omega)\,d\omega
\end{equation}
which is expected to behave smoothly with the various parameters.
We observe that it has a finite limit as
$t_\perp$ goes to $0$ (see Fig.~\ref{moment}). 
This means that, once $t_\perp$ is turned on between the chains, weight
immediately appears predominantly at finite frequencies. 
This typical frequency increases with the strength and with the 
range of the interaction.
This is clearly a signature of some form of incoherent perpendicular transport.

\section{Conclusions}

In this work, different approaches to interchain coherence have been
investigated. As a first step, we have focussed on the energy splitting 
generated by the transverse hopping in the dispersion relation 
of the LL collective modes. By finite size scaling analysis,
we have shown that this splitting was monitored by the LL parameter
$\alpha$.
However, incoherent interchain hopping is found for much smaller
values of $\alpha$ than those predicted by the  RG calculations~\cite{Boies}.
In the second part of this work, we have attempted to make the 
connection between the previous energy splitting and transverse transport
properties. In the regime where $t_\perp$ is still relevant 
($\alpha<\alpha_0$), 
the most important results are that (i) the Drude weight and the
total optical sum rule grow 
less rapidly with $t_\perp$ than in the non interacting case 
and (ii) the Drude weight becomes significantly smaller than
the total sum rule when the intrachain interaction is turned on.
Hence, even when the Drude weight remains finite (when $t_\perp$
is relevant), transverse transport 
is predominantly incoherent in the small $t_\perp$ regime.
How small $t_\perp$ needs to be so that this regime is observed 
depends on the strength of the interaction. Typically, for
$\alpha\sim 0.2$, strong suppression of coherent transport occurs
up to $t_\perp\sim 0.15$. 
This phenomenon could explain
the anomalous transport which is observed experimentally.

\bigskip
We gratefully acknowledge many useful discussions and comments from
C. Hayward and M. Albrecht. {\it Laboratoire de Physique Quantique, Toulouse} is 
{\it Unit\'e Mixte de Recherche CNRS No 5626}. 
We thank IDRIS (Orsay) 
for allocation of CPU time on the C94 and C98 CRAY supercomputers.

%
%


\begin{references}

\bibitem{Anderson_91} P. W. Anderson, Phys. Rev. Lett. {\bf 67}, 3844
(1991); P. W. Anderson, Science {\bf 256}, 1526 (1992).

\bibitem{Cooper} S. L. Cooper and K. E. Gray in {\it Physical Properties of
High Temperature Superconductors IV}, D. M. Ginsberg (ed.) (World
Scientific), 1994; S. L. Cooper et al., Phys. Rev. Lett. {\bf 70}, 1533
(1993); Y. Maeno et al., Nature {\bf 372}, 532 (1994); X.-D. Xiang et al.,
Phys. Rev. Lett. {\bf 68}, 530 (1992); K. Tamasaku, T. Ito, H. Takagi and
S. Uchida, Phys. Rev. Lett. {\bf 72}, 3088 (1994).

\bibitem{Varma} C. M. Varma, P. B. Littlewood, S. Schmitt-Rink, E. Abrahams
and A. E. Ruckenstein, Phys. Rev. Lett {\bf 63}, 1996 (1989).

\bibitem{Haldane} F. D. M. Haldane, J. Phys. C {\bf 14}, 2585 (1981).

\bibitem{jerome} D. J\'erome and H. Schulz,
{\it Adv.  Phys.} {\bf 31}, 299 (1982). 

\bibitem{wzietek} P. Wzietek, F. Creuzet, C. Bourbonnais, D. J\'erome, K.
Bechgaard and P. Batail, J. Phys. (Paris)  I {\bf 3}, 171 (1993).

\bibitem{behnia} K. Behnia et al, Phys. Rev. Lett. {\bf 74}, 5272
(1995).

\bibitem{Boies} D. Boies, C. Bourbonnais, and A.-M.S. Tremblay,
Phys. Rev. Lett. {\bf 74}, 968 (1995).

\bibitem{Anderson_94} D. G. Clarke, S. P. Strong, and P. W. Anderson,
Phys. Rev. Lett. {\bf 72}, 3218 (1994). 

\bibitem{Clarke_96} D. G. Clarke and S. P. Strong, cond-mat/9607141 preprint.

\bibitem{Didier2} F. Mila and D. Poilblanc, Phys. Rev. Lett. {\bf 76}, 287
(1996). 

\bibitem{Fabrizio} M. Fabrizio \& A. Parola, Phys. Rev. Lett. {\bf 70}, 226
(1993). 

\bibitem{Lanczos} C. Lanczos, J. Res. Natl. Bur. Stand. {\bf 45}, 255 (1950).

\bibitem{Didier} D. Poilblanc, H. Endres, F. Mila, M. Zacher, S. Capponi and
W. Hanke, Phys. Rev. B (in press) and cond-mat/9605106 preprint.

\bibitem{Kohn} W. Kohn, Phys. Rev. 133, A 171 (1964).

\bibitem{flux} D. Poilblanc, Phys. Rev. B {\bf 44}, 9562 (1991).

\bibitem{Gros} C. Gros, Z. Phys. B {\bf 86}, 359 (1992).

\bibitem{Voit} J. Voit, Rep. Prog. Phys. {\bf 58}, 977 (1995) and references therein.

\bibitem{Voit93} J. Voit, Phys. Rev. B {\bf 47}, 6740 (1993).

\bibitem{Schulz} H. J. Schulz, {\it ``Correlated Electron Systems''},
p. 199, ed. V. J. Emery (World scientific, Singapore, 1993);
H. J. Schulz, {\it ``Strongly Correlated Electronic Materials:
The Los Alamos Symposium -- 1993''}, 
p. 187, ed. K. S. Bedell, Z. Wang, D. E. Meltzer,
A. V. Balatsky, E. Abrahams (Addison-Wesley, Reading, Massachussets, 1994).

\bibitem{Maldague} P. Maldague, Phys. Rev. B {\bf 16}, 2437 (1977).

\bibitem{note_Convergence} For a finite system of non-interacting fermions,
the $D_{yy}(t_\perp)$ curve is made of linear segments 
the slope of which increases with $t_\perp$. 
The $t_\perp^2$--law is recovered in the 
thermodynamic limit when the average length of the segments vanishes.
This peculiar behavior makes finite size scaling analysis of the transverse
conductivity at finite V difficult. 
\end{references}
\end{document}